\documentclass[useAMS,usenatbib,usegraphicx]{mn2e}
\usepackage{graphicx}
\usepackage{amsmath}
\title[On the origin of the correlations between accretion and line luminosities in PMS stars.]{On the origin of the correlations between the accretion luminosity and emission line luminosities in pre-main sequence stars.}

\author[I. Mendigut\'\i{}a et al.]
{\parbox{\textwidth}{I. Mendigut\'\i{}a$^{1}$\thanks{E-mail: \texttt{I.Mendigutia@leeds.ac.uk}}, 
R.D. Oudmaijer$^{1}$,
E. Rigliaco$^{2}$, 
J.R. Fairlamb$^{1}$, 
N. Calvet$^{3}$, 
J. Muzerolle$^{4}$,  
N. Cunningham$^{1}$ and
S.L. Lumsden$^{1}$
\vspace{0.4cm}}
\\
\parbox{\textwidth}{
$^{1}$School of Physics and Astronomy, University of Leeds, Woodhouse Lane, Leeds LS2 9JT, UK.\\
$^{2}$Swiss Federal Institute of Technology, Department of Physics-Institute for Astronomy, Wolfgang-Pauli-Strasse 27, CH-8093 Zurich, Switzerland\\
$^{3}$Department of Astronomy, University of Michigan, 830 Dennison Building, 500 Church Street, Ann Arbor, MI 48109, USA\\
$^{4}$Space Telescope Science Institute, 3700 San Martin Dr., Baltimore, MD, 21218, USA\\
}}

\begin{document}

\date{MNRAS/Accepted}

\pagerange{\pageref{firstpage}--\pageref{lastpage}} \pubyear{0000}

\maketitle

\label{firstpage}

\begin{abstract}
Correlations between the accretion luminosity and emission line luminosities (L$_{acc}$ and L$_{line}$) of pre-main sequence (PMS) stars have been published for many different spectral lines, which are used to estimate accretion rates. Despite the origin of those correlations is unknown, this could be attributed to direct or indirect physical relations between the emission line formation and the accretion mechanism. This work shows that all (near-UV/optical/near-IR) L$_{acc}$-L$_{line}$ correlations are the result of the fact that the accretion luminosity and the stellar luminosity (L$_{*}$) are correlated, and are not necessarily related with the physical origin of the line. Synthetic and observational data are used to illustrate how the L$_{acc}$-L$_{line}$ correlations depend on the L$_{acc}$-L$_{*}$ relationship. We conclude that because PMS stars show the L$_{acc}$-L$_{*}$ correlation immediately implies that L$_{acc}$ also correlates with the luminosity of all emission lines, for which the L$_{acc}$-L$_{line}$ correlations alone do not prove any physical connection with accretion but can only be used with practical purposes to roughly estimate accretion rates. When looking for correlations with possible physical meaning, we suggest that L$_{acc}$/L$_{*}$ and L$_{line}$/L$_{*}$ should be used instead of L$_{acc}$ and L$_{line}$. Finally, the finding that L$_{acc}$ has a steeper dependence on L$_{*}$ for T-Tauri stars than for intermediate-mass Herbig Ae/Be stars is also discussed. That is explained from the magnetospheric accretion scenario and the different photospheric properties in the near-UV.
\end{abstract}


\section{Introduction}
\label{Sect:intro}

The disk-to-star accretion rate is one of the most important parameters driving the evolution of pre-main sequence (PMS) stars. However, it
is difficult to directly measure the mass accretion rate, for which indirect empirical methods are necessary to estimate it. A widely used method exploits the fact that the accretion luminosity (L$_{acc}$) correlates with the luminosity of various emission lines (L$_{line}$). Despite the unknown origin of these correlations, they are being used to quickly estimate accretion rates. The L$_{acc}$--L$_{line}$ empirical correlations have been derived using samples of PMS stars by comparing their accretion luminosities, mostly obtained from the UV excess and line veiling, with the emission line luminosity \citep[see e.g.][and references therein]{Muzerolle98c,HerczegHillen08,Dahm08,Fang09,Rigliaco12}. Currently, dozens of near-UV -- optical -- near-IR spectral lines have been found to correlate with L$_{acc}$ for classical T Tauri (TT) stars \citep[for instance, the hydrogen Balmer and Paschen series, HeI, OI, NaID and CaII transitions, Br$\gamma$... etc; see e.g.][AL14 hereafter]{Alcala14}. The correlations of the accretion luminosity with several of these lines have been extended both to the sub-stellar and the intermediate-mass Herbig Ae/Be (HAeBe) regimes \citep{Mohanty05,Rigliaco11,Donbrit11,Mendi11,Mendi13a}.

Apart from the observational effort involved to look for additional emission lines that could serve as accretion tracers, several investigations aim to provide physical links between some of the spectral transitions and the accretion process, which would explain the origin of the L$_{acc}$--L$_{line}$ correlations. In a nutshell, either the lines are directly tracing the accreting region \citep[e.g.][]{Muzerolle98a,Muzerolle98b,Kurosawa06,Rigliaco15}, or they trace the accretion indirectly, by probing the accretion-powered outflows and winds \citep[e.g.][]{Hartigan95,Edwards06,Kurosawa11,Kurosawa12}. The correlation with forbidden lines like [OI] (6300 $\AA{}$) exhibited by HAeBes \citep{Mendi11} is more difficult to explain, as this line is not identified with accretion/winds but rather with the surface layers of the circumstellar disks \citep{Acke05}. A further challenge to the various explanations of the origin of the L$_{acc}$--L$_{line}$ correlations is that the variations in the accretion rate as measured from the UV excess do not generally correlate with the observed changes in the line luminosities \citep{Nguyen09,Costigan12,Mendi11,Mendi13a}. However, time delays between different physical processes could be present \citep{Dupree12}.

On the other hand, the accretion luminosity is also found to correlate with the luminosity of the central star (L$_{*}$). The L$_{acc}$--L$_{*}$ correlation extends over $\sim$ 10 orders of magnitude in L$_{acc}$, and $\sim$ 7 orders of magnitude in L$_{*}$, covering all optically visible young stars from the sub-stellar to the HAeBe regime \citep[see e.g.][and references therein]{Natta06,ClarkePringle06,Tilling08,Mendi11,Fairlamb15}. Based on a statistical analysis, \citet{Mendi11} tentatively suggested that the correlation between the accretion luminosity and the luminosity of several emission lines in HAeBe stars could be driven by the common dependence of both luminosities on the stellar luminosity. 

The main goal of this paper is to demonstrate the equivalence of the L$_{acc}$--L$_{line}$ and L$_{acc}$--L$_{*}$ correlations. In
particular, we aim to show that all (near-UV, optical and near-IR) L$_{acc}$--L$_{line}$ correlations in PMS stars are driven by the
relationship between the stellar luminosity and the accretion luminosity, and that therefore the accretion luminosity necessarily
correlates with the luminosity of all spectral lines regardless of their physical origin. Section \ref{Sect:Lacc-Lstar} introduces and partially re-analyses the L$_{acc}$--L$_{*}$ correlation in PMS stars. Section \ref{Sect:formulas} shows the expression that links the L$_{acc}$--L$_{*}$ relationship with the L$_{acc}$--L$_{line}$ correlations. The inter-dependence between both types of correlations is
illustrated in section \ref{Sect:Lacc-Lline_Lacc-Lstar} using both synthetic data and observational data from the literature. Some implications from all the previous analysis are included in section \ref{Sect:consequences}. Finally, section \ref{Sect:conclusion} summarizes our main conclusions.

\section{The L$_{acc}$--L$_*$ correlation}
\label{Sect:Lacc-Lstar}

\begin{figure}
\centering
 \includegraphics[width=8.4cm,clip=true]{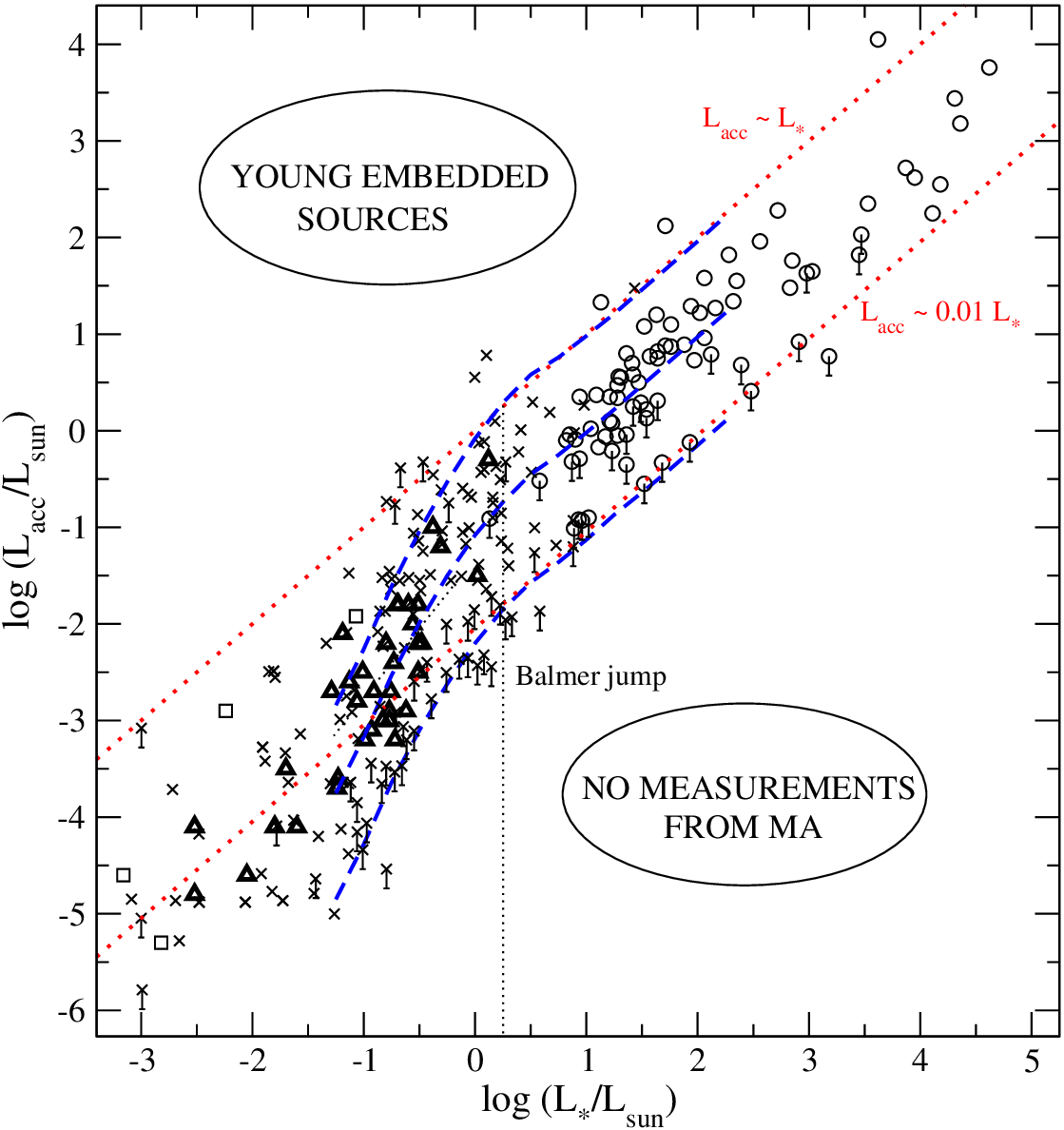}

 \caption{L$_{acc}$--L$_{*}$ correlation for sub-stellar objects and TTs in different star forming regions \citep[crosses; with vertical bars for upper limits;][and references therein]{Natta06,HerczegHillen08}, four (sub-) stellar/planetary companions around PMS stars \citep[squares;][]{Close14,Zhou14}, the Lupus sample from AL14 (dark triangles), and HAeBes \citep[circles; with vertical bars for upper limits;][]{Mendi11,Fairlamb15}. The red diagonal dotted lines indicate L$_{acc}$ = L$_{*}$ and L$_{acc}$ = 0.01L$_{*}$. The three blue diagonal dashed lines represent the accretion luminosities expected from MA modelling for Balmer excesses of 0.70, 0.12 and 0.01 magnitudes (top, mid, and bottom lines, respectively). The vertical dotted line indicates the stellar luminosity at which the Balmer jump becomes apparent in the photospheric spectra (see also Fig. \ref{Figure:excesses_filling}).}
\label{Figure:Lacc-Lstar}
\end{figure}

A representative example of the empirical correlation between the accretion and stellar luminosities \footnote{Its counterpart, the relationship between mass accretion rate and stellar mass, can be derived from the L$_{acc}$--L$_*$ correlation using PMS tracks \citep[see e.g.][]{ClarkePringle06}.} is shown in Fig. \ref{Figure:Lacc-Lstar}. It includes data from the literature for very low-mass TTs and sub-stellar objects/companions (log (L$_{*}$/L$_\odot$) $<$ -1.25), TTs (-1.25 $<$ log (L$_{*}$/L$_\odot$) $<$ 0.75), late-type HAeBes (0.75 $<$ log (L$_{*}$/L$_\odot$) $<$ 2.25), and early type HAeBes (log (L$_{*}$/L$_\odot$) $>$ 2.25). The sources belong to different star forming regions. The graph shows that L$_{acc}$ increases with L$_{*}$, with a relation steeper for the TTs than for the HAeBes. 

According to \citet{ClarkePringle06} and \citet{Tilling08}, the upper bound of the L$_{acc}$--L$_{*}$ correlation (L$_{acc}$ $\sim$ L$_{*}$) is the consequence of sample selection effects; the luminosity of most stars above that limit is dominated by accretion and these objects are in a younger, embedded phase without an optically visible photosphere. The lower bound (L$_{acc}$ $\sim$ 0.01L$_{*}$, mainly for objects with L$_{*}$ $>$ L$_{\odot}$) is limited by accretion detection thresholds (symbols with vertical bars in Fig. \ref{Figure:Lacc-Lstar}). The physical origin of the L$_{acc}$--L$_{*}$ correlation is the subject of active debate. This topic is not analysed here but we refer the reader to several related works \citep[e.g.][]{Padoan05,AlexanderArmitage06,Dullemond06,VorobyovBasu08,Ercolano14}. Instead, our contribution below deals with the observed change in the slope of the L$_{acc}$--L$_{*}$ correlation between the TT and the HAeBe stars \citep{Mendi11,Fairlamb15}.

\begin{figure}
\centering
 \includegraphics[width=8.4cm,clip=true]{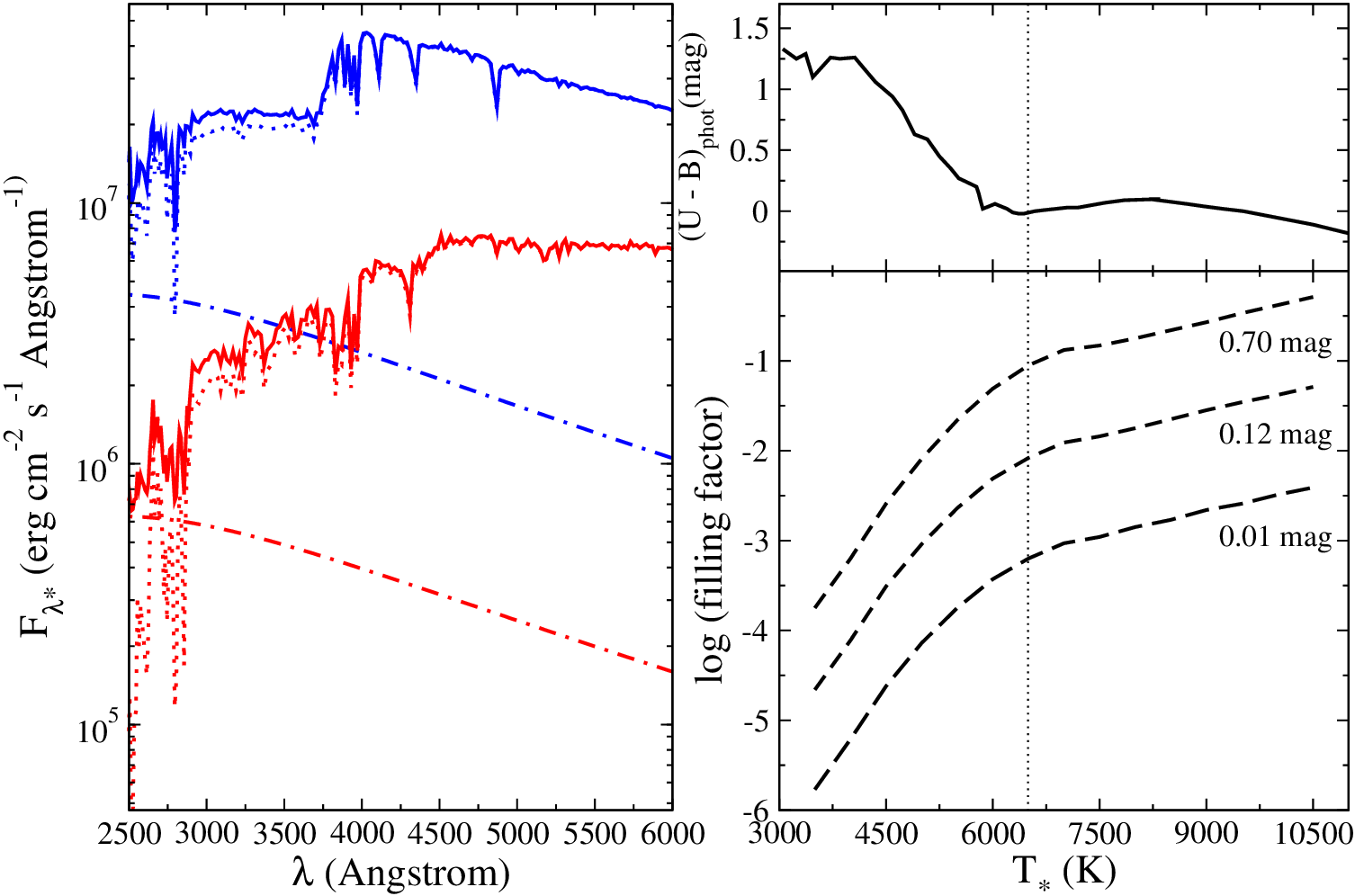}
\caption{Left panel: MA modelling of a typical Balmer excess (0.12 magnitudes) for two representative stars with stellar temperatures of 7500 (blue) and 5500 $K$ (red). The photospheric (Kurucz) spectra, the contribution from accretion, and the total flux obtained from the combination of the previous are represented by the dotted, dot-dashed, and solid lines, respectively. The fluxes are as they would be measured at the stellar surface. Right panels: photospheric $U$ - $B$ colours \citep[taken from][]{KennyonHartmann95} characterizing the Balmer region of the spectrum (top) and filling factors necessary to reproduce a Balmer excess of 0.01, 0.12 and 0.70 magnitudes (bottom) using MA, versus the stellar temperature. The vertical dotted line indicates the stellar temperature at which the Balmer jump becomes apparent in the photospheric spectra (see also Fig. \ref{Figure:Lacc-Lstar}).}
\label{Figure:excesses_filling}
\end{figure}

We constructed a sample of artificial stars representing the TT and HAeBe regime by using synthetic models of
stellar atmospheres \citep{Kurucz93}. The properties of each object are provided in Table \ref{Table:sample1}. Columns two and three show the stellar luminosity and effective temperature. From these, the stellar radii was derived, spanning between 0.7 and 4 R$_\odot$ (column 4). The stellar masses (column 5) were derived assuming log g = 4, and cover the 0.2 -- 6 M$_\odot$ range. Magnetospheric accretion (MA) shock modelling was carried out for each star by adding (blackbody) accretion contributions to the photospheric (Kurucz) spectra \citep[see e.g. the reviews in][]{Calvet00,Mendi13b}. Two representative examples are presented in Fig. \ref{Figure:excesses_filling} (left panel). The shock model was applied following the usual recipes for both the TTs and HAeBes, and we refer the reader to \citet{Calvet98,Mendi11} and \citet{Fairlamb15} for further details. Three different values for the UV excess in the Balmer region of the spectra \citep[from $\sim$ 3500 to 4000 $\rm \AA{}$, as defined in][]{Mendi13a} were modelled for each object assuming typical values for the inward flux of energy carried by the accretion columns (10$^{12}$ erg cm$^2$ s$^{-1}$) and the disk truncation radius (5R$_*$): a ''maximum'' excess (0.70 magnitudes), whose corresponding accretion contribution is L$_{acc}$ $\sim$ L$_{*}$ for L$_{*}$ $\geq$ L$_{\odot}$; a ''minimum'' excess (0.01 magnitudes) representative of the observational limit, and whose corresponding accretion contribution is L$_{acc}$ $\sim$ 0.01L$_{*}$ for L$_{*}$ $\geq$ L$_{\odot}$; and finally, a ''typical'' excess in-between the two previous (0.12 magnitudes). The resulting accretion luminosities are shown in the last three columns of Table \ref{Table:sample1}. These are plotted versus the corresponding L$_{*}$ values (blue diagonal dashed lines in Fig. \ref{Figure:Lacc-Lstar}), matching the overall distribution of data. We note that excesses larger than 0.70 magnitudes could still be measured for the less luminous sources (L$_{*}$ $\leq$ L$_{\odot}$) without reaching the upper bound (L$_{acc}$ $\sim$ L$_{*}$).

\begin{table}
\centering
\renewcommand\tabcolsep{1.5pt}
\caption{Sample of artificial stars. Stellar parameters and accretion luminosities from MA.}
\label{Table:sample1}
\begin{tabular}{rrrrrrrr}
\hline\hline
Star&L$_{*}$&T$_*$&R$_*$&M$_*$&(L$_{acc}$)$_{m}$&(L$_{acc}$)$_{t}$&(L$_{acc}$)$_{M}$ \\
$\sharp$& [log L$_{\odot}$]&(K)&(R$_\odot$)&(M$_\odot$)&[log L$_{\odot}$]&[log L$_{\odot}$]&[log L$_{\odot}$]\\
\hline
1&	-1.25&	3500&	0.65& 0.15&-4.85&-3.75&-2.84\\
2&	-1.00&	4000&	0.66& 0.16&-4.28&-3.17&-2.26\\
3&	-0.75&	4500&	0.70& 0.18&-3.64&-2.53&-1.61\\
4&	-0.50&	5000&	0.75& 0.20&-3.10&-1.99&-1.04\\
5&	-0.25&	5500&	0.83& 0.25&-2.62&-1.50&-0.53\\
6&	0.00&	6000&	0.93& 0.31&-2.20&-1.08&-0.08\\
7&	0.25&	6500&	1.05& 0.40&-1.85&-0.74&0.29 \\
8&	0.50&	7000&	1.21& 0.53&-1.57&-0.45&0.58 \\
9&	0.75&	7500&	1.41& 0.72&-1.36&-0.25&0.76 \\
10&	1.00&	8000&	1.65& 0.99&-1.13&-0.02&0.98 \\
11&	1.25&	8500&	1.95& 1.38&-0.89&0.22 &1.21 \\
12&	1.50&	9000&	2.32& 1.95&-0.64&0.47 &1.46 \\
13&	1.75&	9500&	2.78& 2.80&-0.40&0.72 &1.71 \\
14&	2.00&	10000&	3.34& 4.05&-0.15&0.97 &1.96 \\
15&	2.25&	10500&	4.04& 5.93&0.10 &1.22 &2.22 \\
\hline

\end{tabular}
\begin{minipage}{8.4cm}

  \textbf{Notes.} Columns two to five show the stellar luminosity
  (logarithmic scale, from the integrated Kurucz model atmospheres),
  effective temperature, stellar radius and mass. Columns six to eight
  show the MA accretion luminosities (logarithmic scale) corresponding
  to a minimum (m), typical (t) and maximum (M) Balmer excess of 0.01,
  0.12 and 0.70 magnitudes, respectively.
\end{minipage}
\end{table} 

The fact that the accretion luminosity increases with the stellar luminosity is a natural consequence of MA shock modelling. This is
illustrated in the left panel of Fig. \ref{Figure:excesses_filling}. When the same excess (flux ratio between the solid and dotted lines) is observed in stars of different stellar luminosity, the most luminous stars (blue dotted line) must necessarily have a larger accretion contribution (dot-dashed lines). In order to understand the different slope in the L$_{acc}$ --L$_{*}$ correlation for TT and HAeBe stars, it is important to recall that the accretion contribution, and therefore L$_{acc}$, is proportional to both the temperature of the accretion columns (T$_{col}$) and the filling factor ($f$), which represents the fraction of the stellar surface covered by the accretion shocks. Variations in T$_{col}$ and $f$ move the accretion-generated continuum excess along the wavelength axis and flux axis respectively. The typical value for T$_{col}$ is $\sim$ 10$^4$ $K$  across the TT and HAeBe regimes. Therefore, the excess peaks close to the Balmer region for both types of star. However, their photospheric spectra (i.e. when accretion is not present) are significantly different in that
region. The Balmer jump becomes visible only for stars with log (L$_{*}$/L$_\odot$) $\geq$ 0.25 (i.e. T$_*$ $\geq$ 6500 $K$)\footnote{The Balmer jump disappears again in O stars with T$_*$ $\geq$ 30000 $K$}. This makes the spectra of stars with spectral types F and earlier more similar between them in the Balmer region than for later spectral types. Fig. \ref{Figure:excesses_filling} (top right panel) illustrates the case; the photospheric $U$-$B$ colour characterizing the Balmer region shows a steep dependence on the stellar temperature for cool stars, and flattens for hotter objects. Therefore, in order to reproduce a given Balmer excess, TTs require larger variations in the accretion luminosity than the ones that HAeBes need, for which the slope $\Delta$L$_{acc}$/$\Delta$L$_{*}$ decreases from the TT to the HAeBe regime. The accretion luminosity changes are mainly affected by variations in the filling factor. This is shown in Fig. \ref{Figure:excesses_filling} (bottom right panel), where the filling factors that are needed to reproduce the minimum, typical and maximum model excesses are plotted against the stellar temperature. The change of slope in this panel occurs at the temperature where the Balmer jump appears ($\sim$ 6500 $K$), which corresponds to the stellar luminosity when the slope of the L$_{acc}$ -- L$_{*}$ correlation changes (log (L$_{*}$/L$_\odot$) $\sim$ 0.25). It is noted that we have applied basic MA modelling without considering aspects like the chromospheric contribution to the spectra of TT stars \citep{Manara13} or changes in the disk truncation radius depending on the stellar mass regime \citep{Muzerolle04,Mendi11,Cauley14}. These factors could change the accretion estimates by less than 0.5 dex, without significantly affecting the modelled results in Figs. \ref{Figure:Lacc-Lstar} and \ref{Figure:excesses_filling}.

In summary, the observed difference in the L$_{acc}$--L$_{*}$ correlation between TTs and HAeBes can be explained from the MA scenario and the differences in the near-UV stellar properties between both types of stars. However, we emphasize that the overall L$_{acc}$--L$_{*}$ correlation is not a mere consequence of the MA shock modelling but most probably reflects a deeper physical relationship between both parameters (see e.g. the references at the beginning of this section). For example, the specific slopes shown by different samples in different environments (see e.g the Lupus sample with solid triangles in Fig. \ref{Figure:Lacc-Lstar}) cannot simply be explained from MA. Moreover, the L$_{acc}$--L$_{*}$ correlation seems to arise also in embedded, younger sources, when the accretion luminosities are estimated from a variety of methods (Beltr\'an \& de Wit, to be submitted). Regardless of the underlying physical origin of the L$_{acc}$--L$_{*}$ correlation, for the rest of the paper it will be enough to remind that this arises whenever a significant sample of PMS stars is considered.

\section{The accretion-stellar-line luminosity relation}
\label{Sect:formulas} 
The relation between the accretion and stellar luminosities is usually expressed in the literature as L$_{acc}$ $\propto$ L$_{*}^b$. This can also be written as a linear expression, which is a reasonable approach when the TT and HAeBe regimes are studied separately. For a given star, we will assume that L$_{acc}$ and L$_{*}$ can then be related by:
\begin{equation}
\label{Eq.Lacc_Lstar}
\log \left(\frac{L_{acc}}{L_{\sun}}\right) = a + b \times \log \left(\frac{L_*}{L_{\sun}}\right),
\end{equation}
with $a$ and $b$ constants that depend on the star considered. When a sample of stars is studied, $a$ and $b$ represent the intercept and
the slope of a linear fit to the data. This situation will be analysed in section \ref{Sect:Lacc-Lline_Lacc-Lstar}.

The luminosity of a spectral line can be computed by multiplying the line equivalent width (EW) and the luminosity (per unit wavelength) of the adjacent continuum (L$_{\lambda}^{c}$): 

\begin{equation}
\label{Eq.Lline}
L_{line} = L_{\lambda}^{c} \times EW = \left(\frac{\alpha \cdot EW}{\beta}\right) \times L_*,
\end{equation}
with $\alpha$ the (dimensionless) excess of the (dereddened) continuum with respect to the photosphere at the wavelength of the line ($\alpha$ = $L_{\lambda}^{c}$/$L_{\lambda *}$ $\gid$ 1), and $\beta$ the ratio between the total stellar luminosity and the stellar
luminosity at that wavelength ($\beta$ = L$_{*}$/L$_{\lambda *}$ $>>$ 1, in units of wavelength). The stellar luminosity in the second term
of Eq. \ref{Eq.Lline} was introduced in Eq. \ref{Eq.Lacc_Lstar}, obtaining:
\begin{equation}
\label{Eq.Lacc_Lline}
\log \left(\frac{L_{acc}}{L_{\sun}}\right) = A + B \times \log \left(\frac{L_{line}}{L_{\sun}}\right),
\end{equation}
which is again a linear expression, with:
\begin{equation}
\label{Eq.AB}
\begin{split}
&A = a - b \times \log \left(\frac{L_{line}}{L_*}\right); \\ 
&B = b,
\end{split}
\end{equation}
where L$_{line}$/L$_*$ = $\alpha$EW/$\beta$, is the line to stellar luminosity ratio. Therefore, if the accretion luminosity of a given star can be derived from its stellar luminosity through Eq. \ref{Eq.Lacc_Lstar}, then the same accretion luminosity can be recovered from the luminosity of any emission line through Eqs. \ref{Eq.Lacc_Lline} and \ref{Eq.AB}, with $A$ and $B$ constants that depend on the star and the line considered. Equations \ref{Eq.Lacc_Lstar} and \ref{Eq.Lacc_Lline} are equivalent because both express a common dependence of the accretion luminosity on the stellar luminosity (Eq. \ref{Eq.Lline}).

\section{The dependence of the L$_{acc}$--L$_{line}$ correlations on the L$_{acc}$--L$_*$ relation}
\label{Sect:Lacc-Lline_Lacc-Lstar}
In this section we use both synthetic and empirical data to illustrate
the dependence of the L$_{acc}$--L$_{line}$ correlations on the
L$_{acc}$--L$_*$ relation. Our first analysis provides a simple
qualitative example on how the shape of the L$_{acc}$--L$_*$
relationship has a strong effect on the L$_{acc}$--L$_{line}$
correlations. We use the sample of artificial stars introduced in the
previous section (see the first five columns of
Table \ref{Table:sample1}). The Kurucz models were used to
calculate L$_{\lambda}^{c}$ and $\beta$ at 6000 $\rm \AA{}$, whose
values are presented in columns two and three of Table
\ref{Table:sample2}. Random EWs (between 1 and 10 $\rm \AA{}$,
column four) are assigned to each object. These range in EW is representative of emission lines with intermediate strength such as the Ca II or OI transitions. The luminosity of an artificial emission line at
6000 $\rm \AA{}$ (column five) can then be obtained 
from Eqs. \ref{Eq.Lline}.  

\begin{table}
\centering
\renewcommand\tabcolsep{3.2pt}
\caption{Sample of artificial stars. Continuum and line properties.}
\label{Table:sample2}
\begin{tabular}{rrrrr}
\hline\hline
Star&L$_{6000}^{c}$&$\beta$($\lambda$ = 6000 $\rm \AA{}$)&EW& L$_{6000}$\\
$\sharp$&[log L$_{\odot}$ $\rm \AA{}$] &($\rm \AA{}$)&($\rm \AA{}$)& [log L$_{\odot}$]\\
\hline
1& -5.64&   24317&  7 &-4.79\\
2& -5.13&   13363&  5 &-4.43\\
3& -4.74&   9790&   2 &-4.44\\
4& -4.42&   8343&   9 &-3.47\\
5& -4.14&   7792&   1 &-4.14\\
6& -3.88&   7573&   2 &-3.58\\
7& -3.63&   7587&   4 &-3.03\\
8& -3.39&   7698&   7 &-2.54\\
9& -3.15&   7880&   10&-2.15\\
10&-2.92&   8284&   3 &-2.44\\
11&-2.69&   8773&   6 &-1.91\\
12&-2.48&   9603&   8 &-1.58\\
13&-2.27&   10581&  4 &-1.67\\
14&-2.07&  11723&  1 &-2.07\\
15&-1.86&  12905&  6 &-1.08\\

\hline

\end{tabular}
\begin{minipage}{8.4cm}

  \textbf{Notes.} Columns two to five list the luminosity of the
  continuum at 6000 $\rm \AA{}$ (logarithmic scale), the ratio between the
  total, star + accretion, luminosity and the stellar luminosity at
  6000 $\AA{}$, a random EW of an hypothetical emission line assigned
  to each star (between 1 and 10 $\rm \AA{}$), and its corresponding
  luminosity at 6000 $\rm \AA{}$ (logarithmic scale).
\end{minipage}
\end{table} 

\begin{figure}
\centering
 \includegraphics[width=8.4cm,clip=true]{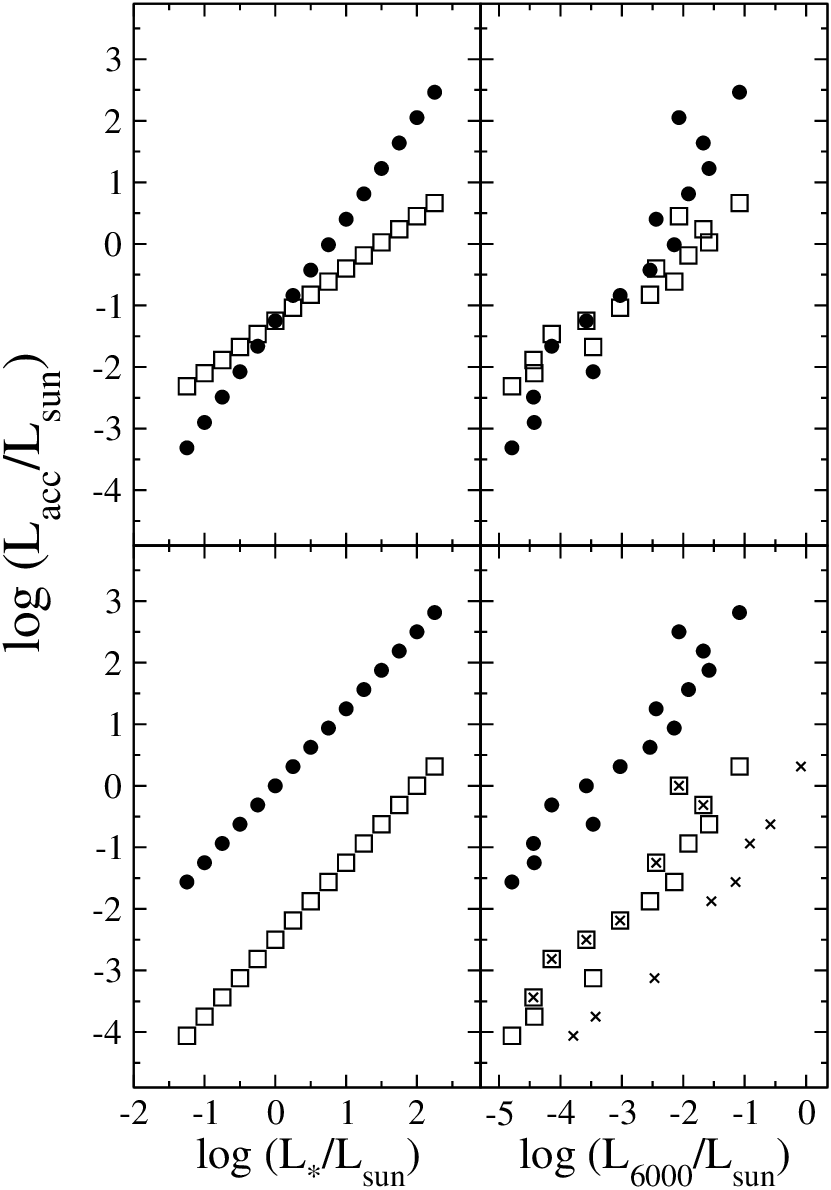}
\caption{Results for the  sample of artificial stars, showing how
  changes in the assumed L$_{acc}$--L$_{*}$ correlation (left panels)
  have an effect on the L$_{acc}$--L$_{line}$ relation (right panels) by
  changing the slope (top panels) and the intercept (bottom panels) of
  the former correlations. The crosses in the bottom-right panel
  represent the line luminosities obtained from a wider EW range,
  when the EWs $\geq$ 5 \AA{} in column 5 of Table
  \ref{Table:sample2} are multiplied by a factor 10.}
\label{Figure:slope_intercept}
\end{figure}

The top left panel of Figure \ref{Figure:slope_intercept} shows two
different L$_{acc}$--L$_{*}$ linear relations assumed for the
sample. Both have the same intercept but a different slope. The
reverse is shown in the bottom left panel, in which the slope is kept
constant and the intercept varies. The right hand panels show the
corresponding accretion luminosities versus the luminosity of the
artificial line at 6000 $\rm \AA{}$. The L$_{acc}$--L$_{line}$
correlations follow the changes introduced in the L$_{acc}$--L$_{*}$
relation, varying their slopes and intercepts. The range in the EW
used  only affects the scatter of
the L$_{acc}$--L$_{line}$ correlation, but this is ultimately
determined by the L$_{acc}$--L$_{*}$ relation. As introduced in
section \ref{Sect:formulas},  the
contribution of the continuum to the line luminosity dominates over
the EW, and both the continuum and the accretion luminosities are correlated with the stellar luminosity. In order to illustrate this, the EW
range was increased multiplying by 10 all the EWs $\geq$ 5 \AA{} in
column 4 of Table \ref{Table:sample2}, and keeping
the rest unmodified. This range in EW is representative of a strong emission line such as H$\alpha$. The new line luminosities are plotted with
crosses in the bottom-right panel of
Fig. \ref{Figure:slope_intercept}, showing that for wider (narrower)
EW ranges, the scatter in the L$_{acc}$--L$_{line}$ correlation
increases (decreases), but the correlation remains.

Before using real data from the literature to illustrate how the
L$_{acc}$--L$_{line}$ empirical correlations are driven by the
L$_{acc}$--L$_*$ relation, the equations described in the previous
section have to be slightly modified. There, the values of
L$_{acc}$ were given by Eqs. \ref{Eq.Lacc_Lstar} and
\ref{Eq.Lacc_Lline}, where $a$, $b$, $A$ and $B$ differ depending on
the individual star and spectral line. In practice, the values for the
slopes and intercepts of these equations are estimated using
linear regression fitting, which provide unique $a$ and $b$ values for a given sample
of stars, as well as unique $A$ and $B$ values for a given spectral
line. In this case it can be shown (see Appendix \ref{appendix}) that
the slopes and intercepts of the L$_{acc}$--L$_*$ and the
L$_{acc}$--L$_{line}$ empirical correlations are related by:

\begin{equation}
\label{Eq.AB_mod}
\begin{split}
&A \sim a - b \times \epsilon \times \left<\log \frac{L_{line}}{L_*}\right>; \\ 
&B = b \times \epsilon; \\
&\epsilon = \frac{r_{line} \times \sigma_{*}}{r_{*} \times \sigma_{line}} \sim 1,\\
\end{split}
\end{equation}
where $A$, $a$; $B$, and $b$ represent the intercepts and slopes of the
L$_{acc}$--L$_*$ and L$_{acc}$--L$_{line}$ correlations, as derived
from least squares linear regression fitting, $<$log
L$_{line}$/L$_*$$>$ the mean (logarithmic) line to stellar luminosity
ratio, r$_{*}$ and r$_{line}$ the correlation coefficients of the
L$_{acc}$--L$_*$ and L$_{acc}$--L$_{line}$ linear fits, and
$\sigma$$_{*}$ and $\sigma$$_{line}$ the standard deviations of the
log (L$_*$/L$_{\odot}$) and log (L$_{line}$/L$_{\odot}$) values.

In short, when the empirical L$_{acc}$--L$_*$ and
L$_{acc}$--L$_{line}$ correlations are compared, Eqs. \ref{Eq.AB_mod}
should be used instead of Eqs. \ref{Eq.AB}. These are slightly
modified by including the $\epsilon$ parameter, which accounts for the
fact that the empirical correlations are in practice derived from
(least-squares) linear fitting\footnote{Linear regression fits
  obtained from methods different than the usual least-squares are not
  considered in this work. The $\epsilon$ parameter should be
  eventually modified if other linear regression methods are used.}.
      
We use the observational data in AL14 to illustrate the dependence of
the L$_{acc}$--L$_{line}$ empirical correlations on the
L$_{acc}$--L$_*$ relation. These authors studied a sample of 36
low-mass TTs in the Lupus star forming region, for which they derived
stellar parameters, accretion rates from the UV excess, and
L$_{acc}$--L$_{line}$ empirical correlations for dozens of emission
lines in the spectral range from the near-UV to the near-infrared. To
our knowledge, this work contains the largest number of spectral
lines for which this type of correlations are
derived. Another advantage is that for each star the accretion
luminosity and the luminosity of all spectral lines were derived from
the same spectrum, avoiding the problem of variability. In addition,
all the stars are located at a similar distance, which guarantees that
the correlations were not artificially stretched when the fluxes are
multiplied by the squared distances to derive the (accretion and line)
luminosities. Therefore, we consider the L$_{acc}$--L$_*$ and
L$_{acc}$--L$_{line}$ correlations in AL14 as representative for
similar correlations provided in the literature (see e.g the
references in section \ref{Sect:intro}).

\begin{figure}
\centering
 \includegraphics[height=200mm,clip=true]{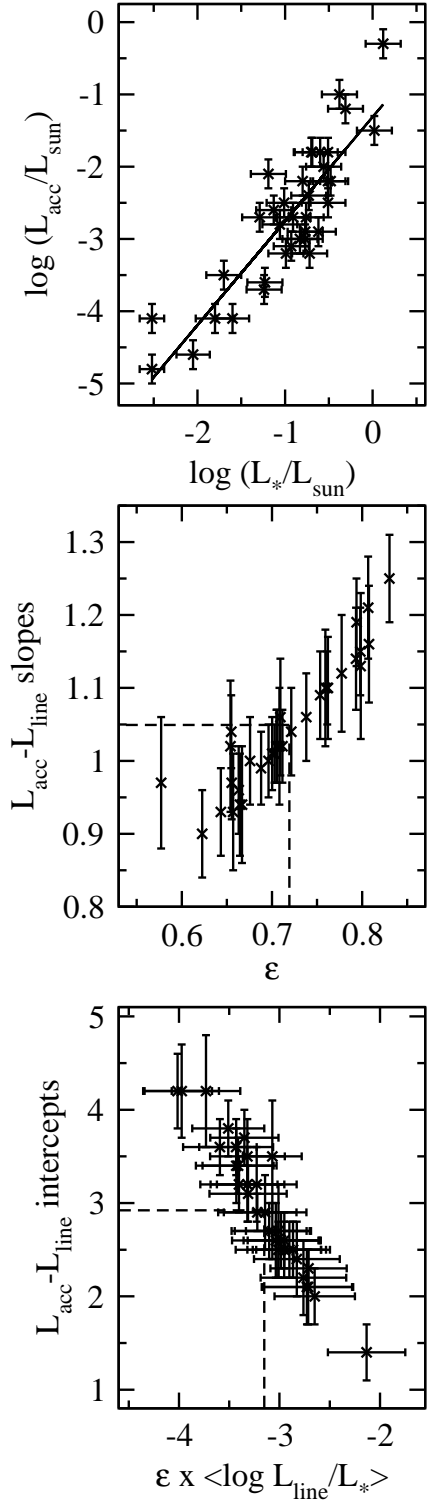}
\caption{Based on results in AL14. Top panel: Accretion versus stellar
  luminosity. The best linear fit is log (L$_{acc}$/L$_\odot$) $\sim$
  -1.3 + 1.4$\times$log (L$_*$/L$_\odot$) (solid line). Mid and bottom
  panels: Slopes and intercepts of the L$_{acc}$--L$_{line}$ empirical
  correlations versus the $\epsilon$ parameter (mid-panel) and the
  $\epsilon$-modified mean (logarithmic) line to stellar luminosity
  ratio (bottom-panel). The dashed lines indicate the mean values for
  the $x$ and $y$ axis, related from the slope and intercept of the top panel correlation by: $<$y$>$ = 1.4$<$x$>$ (mid panel), and $<$y$>$ = -1.3 - 1.4$<$x$>$ (bottom panel).}
\label{Figure:comp_alcala}
\end{figure}

The top panel of Fig. \ref{Figure:comp_alcala} shows the accretion and
stellar luminosities of the stars studied by AL14. The observed trend
is best fitted by log (L$_{acc}$/L$_\odot$) $\sim$ -1.3 +
1.4$\times$log (L$_*$/L$_\odot$) (solid line).
The slopes and intercepts of the L$_{acc}$--L$_{line}$ empirical
correlations derived by AL14 (see their table 4), which are
exactly recovered by Eqs. \ref{Eq.AB_mod}, are plotted in the mid and
bottom panels of Fig. \ref{Figure:comp_alcala} versus $\epsilon$ and
$\epsilon$ $\times$ $<$log L$_{line}$/L$_*$$>$, respectively. The mid
panel shows that the slopes of the L$_{acc}$--L$_{line}$ empirical
correlations are a factor $\epsilon$ smaller than the slope of the
L$_{acc}$--L$_{*}$ correlation shown by the sample. As
expected from Eqs. \ref{Eq.AB_mod}, the L$_{acc}$--L$_{line}$
empirical correlations become steeper when $\epsilon$ increases,
eventually reaching a slope of $\sim$ 1.4 for $\epsilon$ = 1. The
bottom panel shows the expected linear decrease of the intercepts of
the L$_{acc}$--L$_{line}$ correlations with the ($\epsilon$-modified)
line to stellar luminosity ratio. Equations \ref{Eq.AB_mod} also
imply that the typical (mean) slope of all L$_{acc}$--L$_{line}$
correlations is given by the slope of the L$_{acc}$--L$_{*}$
  correlation of the sample, corrected by the mean value of
  $\epsilon$; $<$B$>$ = $b$ $\times$ $<$$\epsilon$$>$. Similarly, it
can be derived that the mean intercept of the L$_{acc}$--L$_{line}$
correlations is given by $<$A$>$ = $a$ - $b$ $\times$ $<$$\epsilon$
$\times$ $<$log L$_{line}$/L$_*$$>$$>$. The two previous relations are
also observed in the AL14 data, the mean values indicated with the
dashed lines perpendicular to both axis in the mid and bottom panels
of Fig \ref{Eq.AB_mod}.

In summary, the analysis of both a sample of artificial stars and
representative empirical data shows that the L$_{acc}$--L$_{line}$
correlations are driven by the underlying L$_{acc}$--L$_*$ relation shown by the sample of stars under study.

\section{Consequences}
\label{Sect:consequences}
The first consequence of the analysis in the previous sections is that
the fact that PMS stars show the L$_{acc}$--L$_{*}$ correlation
immediately implies that L$_{acc}$ also correlates with the luminosity
of any (near-UV-optical-near-UR) emission line, regardless of
the physical origin of the spectral transition.
Indeed, it even correlates with the luminosity of a randomly general
artificial emission line (right panels of Fig.
\ref{Figure:slope_intercept}). As mentioned earlier, the scatter of
the L$_{acc}$--L$_{line}$ correlations increases when the lines' EWs
exhibit a larger range. A similar effect occurs for stars with strong
excess at short, UV, wavelengths and long, IR, wavelengths. For
lines observed a these short and long wavelengths, the ratio
$\alpha$EW/$\beta$ (i.e. the line to stellar luminosity ratio;
Eq. \ref{Eq.Lline}) becomes significant, which could make the
L$_{acc}$--L$_{line}$ correlations much more scattered or eventually
disappear.

For the other lines, the L$_{acc}$--L$_{line}$ correlations are mainly
determined by the L$_{acc}$--L$_{*}$ dependence shown by the sample 
under analysis. The intercepts and slopes provided in the literature
for the L$_{acc}$--L$_{*}$ correlation ($a$ and $b$ in
Eq. \ref{Eq.Lacc_Lstar}) vary depending on the sample of stars
considered \citep[][and references therein]{Fairlamb15}. Based on
those works, a conservative observational limit is -2.5 $\leq$ $a$
$\leq$ 0, 0.8 $\leq$ $b$ $\leq$ 2. Consequently (see Eqs. \ref{Eq.AB}
and \ref{Eq.AB_mod}), the slopes of all L$_{acc}$--L$_{line}$
empirical correlations should also range in between $\sim$ 0.8 and 2,
whereas the intercepts should all be $>$ 0 and decrease as the mean
line to stellar luminosity ratio increases. These predictions agree
with all L$_{acc}$--L$_{line}$ published correlations based on
observational data, to our knowledge. Interestingly, if two samples of
stars show a different slope in their corresponding L$_{acc}$--L$_*$
correlations, then the slopes of the L$_{acc}$--L$_{line}$ ones are
simply related via $B$' $\sim$ $B$ $\times$ ($b$'/$b$) (assuming that
the $\epsilon$ factors in Eq. \ref{Eq.AB_mod} are roughly similar in
both samples). This effect has already been  observed. \citet{Mendi11}
reported a slight decrease in the slope of the L$_{acc}$--L$_*$
correlation of a sample of 34 HAeBe stars with respect to TTs
\citep[see also Fig. \ref{Figure:Lacc-Lstar} and][]{Fairlamb15}. As
discussed there, the slopes of the L$_{acc}$--L$_{line}$ empirical
correlations for the three lines studied (H$\alpha$, [OI] (6300
\AA{}), and Br$\gamma$) also show a similar decrease.

That L$_{acc}$ correlates with L$_{line}$ is ultimately due
to a common dependence of both luminosities on the stellar
brightness. Because of this and the reasons above, the
L$_{acc}$--L$_{line}$ correlations alone cannot be seen as proof for
either a direct or indirect physical connection between the spectral
transitions and the accretion process. However, they are still
useful expressions that can be applied to easily derive accretion
luminosities without the need for sophisticated modelling of the UV
excess. A basic measurement of a line luminosity suffices. Given that both observational L$_{acc}$--L$_{line}$ and L$_{acc}$--L$_{*}$ correlations show a roughly similar scatter (around $\pm$ 1 dex in L$_{acc}$), the latter can also be used to easily derive accretion rates from the stellar luminosity. 

Analogously, since L$_{line}$ necessarily correlates with L$_*$
(Eq. \ref{Eq.Lline}), correlations between L$_{line}$ and L$_*$ alone
can not be taken as a possible physical link between the spectral
transition and the stellar luminosity \citep[see also][]{Natta14}. By extension, the luminosities
of two different emission lines should also correlate with each other
because of the common dependence on the stellar luminosity. Again,
exceptions are possible for lines at short/long wavelengths in stars
with strong excesses \citep[see e.g.][]{Meeus12}.

\begin{figure}
\centering
 \includegraphics[width=8.4cm,clip=true]{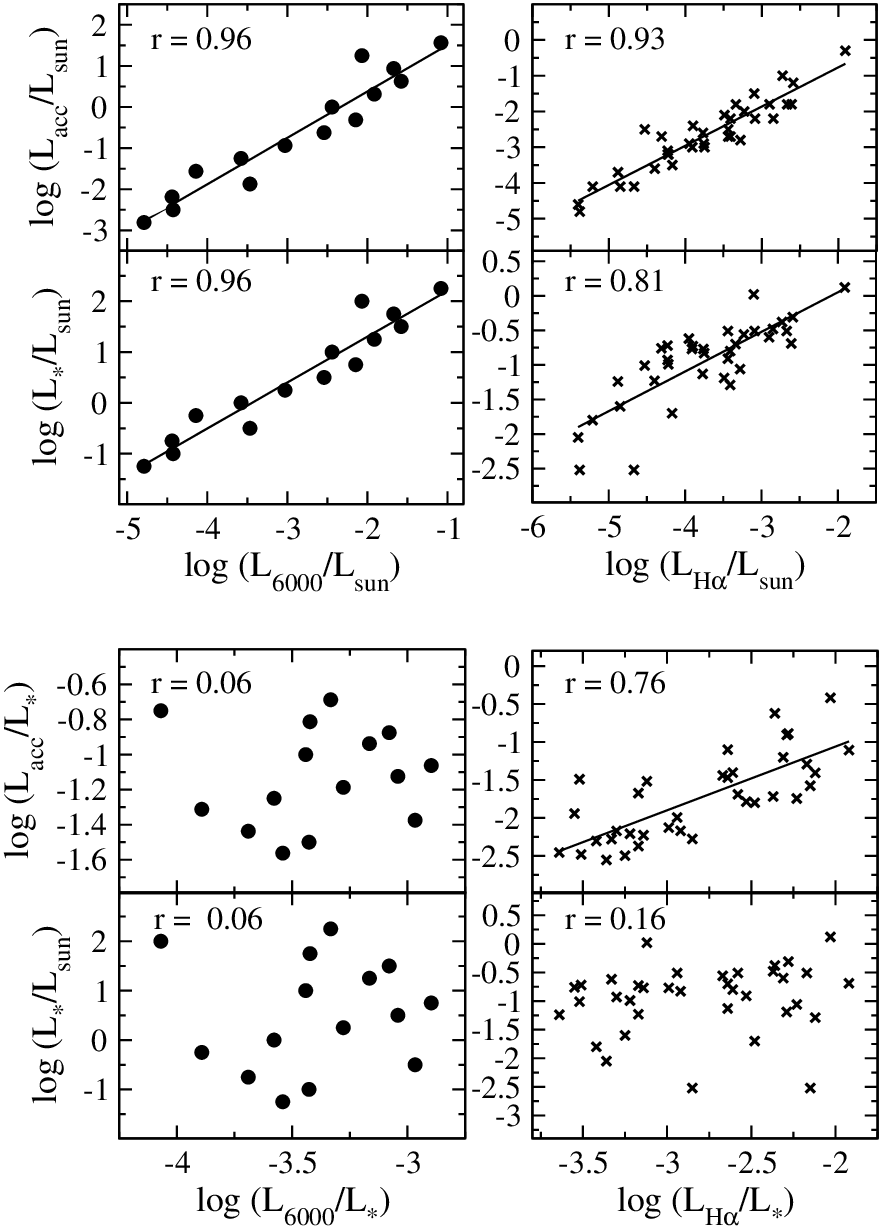}
\caption{Comparison between different luminosities normalized by the
  solar and the stellar luminosity, as indicated in the axes' labels. The left
  panels refer to the  sample of artificial stars in Table
  \ref{Table:sample2}, and the right panels to real observations from
  AL14. Linear regression fits are overplotted for those cases with
  large enough correlation coefficients (r $>$ 0.50; r-values
  indicated in each panel).}
\label{Figure:correct_incorrect}
\end{figure}

In order to infer from correlations possible physical links involving the luminosity of a spectral line or the accretion luminosity, it is necessary to get rid of the common dependence of both parameters on the stellar luminosity. This can be done by
dividing L$_{line}$ and L$_{acc}$ by L$_*$. Fig. \ref{Figure:correct_incorrect} (top  panels) shows the
L$_{acc}$--L$_{line}$ correlation for the sample of artificial stars from
Table \ref{Table:sample2} and a given L$_{acc}$--L$_*$ relation, and the intrinsic correlation between
the stellar and line luminosities. However, the bottom left panels show that both L$_{acc}$/L$_*$ and L$_*$ do not
correlate with L$_{line}$/L$_*$, as expected from an artificial line
created with random EWs. The right panels of the same figure show the
results of the same exercise using real data from AL14. As expected,
the H$\alpha$ luminosity correlates with both the accretion and 
stellar luminosities, which as we have discussed has no possible physical interpretation.
In contrast with the previous example, in this case the H$\alpha$ line to stellar
luminosity ratio is still correlated with the accretion to stellar
luminosity ratio but not with the stellar luminosity itself,
supporting the idea that this line is mainly
driven by accretion and not by the stellar brightness.

With this perspective in mind, we have confirmed that all line
luminosities provided in AL14 correlate with each other, as
expected. We also have checked that when the line luminosities are normalized by the
stellar luminosities, some correlations remain while others disappear,
indicating the presence or absence of a physical link between the
different spectral transitions. For example, for H$\alpha$ and
Br$\gamma$ the correlation is not only between their line
luminosities but also between their line to
stellar luminosity ratios, suggesting a common physical
origin for both transitions. In contrast, despite the fact that the luminosities of the HeII (4686
$\rm \AA{}$) and the CaII (8498 $\rm \AA{}$) lines correlate, their line to
stellar luminosities do not show a significant correlation, suggesting
a different physical origin. 

Finally, when the general L$_{acc}$ -- L$_{*}$ correlation analysed in
section \ref{Sect:Lacc-Lstar} is transformed into
L$_{acc}$/L$_{*}$ vs L$_{*}$, no trend is shown either for the whole
sample or for specific samples like the Lupus objects in AL14. The
vast majority of the objects have 0.01 $\leq$ $L_{acc}$/L$_{*}$ $\leq$
1 (diagonal dotted lines in Fig. \ref{Figure:Lacc-Lstar}) for all
stellar luminosity bins. The typical value of $L_{acc}$/L$_{*}$ is
0.1, which corresponds to the modelled, typical Balmer excess of 0.12
magnitudes. For the less luminous sources (L$_*$ $<$ L$_\odot$),
smaller L$_{acc}$/L$_{*}$ ratios can still be obtained from the same
Balmer excess detection limit. As discussed in Sect. \ref{Sect:Lacc-Lstar}, this is the expected
consequence of the MA scenario and the photospheric properties of the
stars in the near-UV.

It is beyond the scope of this work to carry out a detailed study on
physical correlations involving stellar, line, and accretion luminosities. Instead, we have provided
several examples to suggest that correlation analysis aiming to infer
physical consequences should use L$_{line}$/L$_*$ and L$_{acc}$/L$_*$
and not simply L$_{line}$ and L$_{acc}$.
  
\section{Summary and conclusions}
\label{Sect:conclusion}

The L$_{acc}$--L$_{*}$ empirical correlation in PMS stars has been
partially re-analysed taking into account the newly available accretion
rates for HAeBes. Despite the physical origin of the L$_{acc}$--L$_{*}$ correlation remains
  subject to debate, the observed change of slope from the TT to the HAeBe regime can be understood from the MA scenario and the near-UV
  photospheric properties of the stars.  

We have shown that the fact that PMS stars show the L$_{acc}$--L$_{*}$
correlation immediately implies that L$_{acc}$ also correlates with
the luminosity of any (near-UV, optical, near-IR) emission line,
regardless of the physical origin of the spectral transition. The
overall L$_{acc}$--L$_{line}$ trends are mainly governed by the
L$_{acc}$--L$_{*}$ correlation shown by the sample of stars
under analysis. In particular, the slopes of the L$_{acc}$-L$_{line}$
empirical correlations should typically be between $\sim$ 0.8 and 2
for all spectral lines, which are the observational limits for the
slope of the L$_{acc}$-L$_{*}$ relation. The intercepts also depend on
the L$_{acc}$--L$_{*}$ correlation,  all of which are $>$ 0 and
increasing as the line to stellar luminosity ratio decreases.

Despite the fact that the L$_{acc}$--L$_{line}$ correlations alone do
not constitute an indication of any direct or indirect physical link
between the spectral transitions and accretion, they are a useful tool
to easily derive estimates of the accretion rates. The
L$_{acc}$--L$_{*}$ correlations can be used for the same purpose. Similarly,
correlations between stellar and line luminosities, or between
different line luminosities, do not indicate a physical
relation between the parameters involved. Instead, we suggest that the line to
stellar and accretion to stellar luminosity ratios should be used when
investigating the possible physical origin of the various correlations.

\section*{Acknowledgments}
The authors sincerely acknowledge A. Natta, W.J. de Wit and
M. Beltr\'an for the fruitful discussions that have served to improve
the contents of this manuscript, as well as the anonymous referee
for her/his useful comments.\\

\appendix
\section{Relation between the L$_{acc}$--L$_{line}$ and L$_{acc}$--L$_*$ linear regression correlations}
\label{appendix}
Consider a sample of $N$ stars for which measurements of accretion and
stellar luminosities [log (L$_{acc}$/L$_{\odot}$)$_1$,..., log
  (L$_{acc}$/L$_{\odot}$)$_N$; log (L$_{*}$/L$_{\odot}$)$_1$,..., log
  (L$_{*}$/L$_{\odot}$)$_N$)] are available. A linear fit to the data provides an
expression that links both variables through
\begin{equation}
\label{Eq.Lacc_Lstar_ap}
\log \left(\frac{L_{acc}}{L_{\sun}}\right) = a + b \times \log \left(\frac{L_*}{L_{\sun}}\right),
\end{equation}
with $a$ and $b$ constants representing the intercept and the slope, which from least-squares linear regression are given by
\begin{equation}
\label{Eq.ab_ap}
\begin{split}
&b = r_{*} \times \left(\frac{\sigma_{acc}}{\sigma_{*}}\right); \\ 
&a = \left<\log \left(\frac{L_{acc}}{L_{\odot}}\right)\right> - b \times \left<\log \left(\frac{L_{*}}{L_{\odot}}\right)\right>, 
\end{split}
\end{equation} 
where r$_{*}$ is the correlation coefficient ($\sim$ 1 for well correlated data), and $\sigma$$_{acc}$, $\sigma$$_{*}$; $<$log L$_{acc}$/L$_\odot$$>$, and $<$log L$_{*}$/L$_\odot$$>$ the standard deviations and the means of the log (L$_{acc}$/L$_{\odot}$)$_i$ and log (L$_{*}$/L$_{\odot}$)$_i$ values, respectively.

Similarly, if for the same sample of stars there are additional measurements of the luminosity of a given emission line [log (L$_{line}$/L$_{\odot}$)$_1$,..., log (L$_{line}$/L$_{\odot}$)$_N$], then a linear fit provides 
\begin{equation}
\label{Eq.Lacc_Lline_ap}
\log \left(\frac{L_{acc}}{L_{\sun}}\right) = A + B \times \log \left(\frac{L_{line}}{L_{\sun}}\right),
\end{equation}
with $A$ and $B$ constants given by least-squares linear regression
\begin{equation}
\label{Eq.AB_ap}
\begin{split}
&B = r_{line} \times \left(\frac{\sigma_{acc}}{\sigma_{line}}\right); \\ 
&A = \left<\log \left(\frac{L_{acc}}{L_{\odot}}\right)\right> - B \times \left<\log \left(\frac{L_{line}}{L_{\odot}}\right)\right>, 
\end{split}
\end{equation}
where the correlation coefficient, standard deviations, and means now refer to the [log (L$_{acc}$/L$_{\odot}$)$_i$, log (L$_{line}$/L$_{\odot}$)$_i$] values.

The standard deviation $\sigma$$_{acc}$ can be found in the expression for $b$ of Eq. \ref{Eq.ab_ap}, and then introduced in the expression for $B$ of Eq. \ref{Eq.AB_ap}, providing the expression relating the slopes of the L$_{acc}$ -- L$_{*}$ and L$_{acc}$ -- L$_{line}$ linear correlations:
\begin{equation}
\label{Eq.B_b_ap}
\begin{split}
&B = \epsilon \times b; \\ 
&\epsilon = \frac{r_{line} \times \sigma_{*}}{r_{*} \times \sigma_{line}}. 
\end{split}
\end{equation}

On the other hand, the mean value $<$log L$_{acc}$/L$_\odot$$>$ can be found in the expression for $a$ of Eq. \ref{Eq.ab_ap}, and introduced in the expression for $A$ of Eq. \ref{Eq.AB_ap}. Also considering Eq. \ref{Eq.B_b_ap}, the expression that relates both intercepts is:
\begin{equation}
\label{Eq.A_a_ap}
A = a - b \times \epsilon \times \left[\left<\log \left(\frac{L_{line}}{L_{*}}\right)\right> - \left(\frac{1 - \epsilon}{\epsilon}\right) \times \left<\log \left(\frac{L_{*}}{L_{\odot}}\right)\right>\right].
\end{equation}  
The third term could been neglected ((1 - $\epsilon$)/$\epsilon$ $\sim$ 0) compared with the two other terms in the previous equation.

\label{lastpage}

\begin{thebibliography}{99}
\bibitem[Acke et al.(2005)]{Acke05} Acke, B., van den Ancker, M.E., Dullemond, C.P. 2005, A\&A, 436, 209
\bibitem[Alcal\'a et al.(2014)]{Alcala14} (AL14) Alcal\'a, J.M.; Natta, A.; Manara, C.F. et al. 2014, A\&A, 561, A2
\bibitem[Alexander \& Armitage(2006)]{AlexanderArmitage06} Alexander R.D. \& Armitage P.J. 2006, ApJ, 639, L83
\bibitem[Calvet \& Gullbring(1998)]{Calvet98} Calvet, N., \& Gullbring, E. 1998, ApJ, 509, 802
\bibitem[Calvet et al.(2000)]{Calvet00} Calvet, N.; Hartmann, L.; Strom, S.E. 2000, Evolution of Disk Accretion, in Protostars and Planets IV, ed. V. Mannings, A.P. Boss, \& S.S. Russell (University of Arizona Press, Tucson), 377
\bibitem[Cauley \& Johns-Krull(2014)]{Cauley14} Cauley, P.W. \& Johns-Krull, C.M. 2014, ApJ, 797, 112
\bibitem[Clarke \& Pringle(2006)]{ClarkePringle06} Clarke, C.J., Pringle, J.E. 2006, MNRAS, 370, L10
\bibitem[Close et al.(2014)]{Close14} Close, L.M.; Follette, K.B.; Males, J.R. et al. 2014, ApLJ, 781, L30 
\bibitem[Costigan et al.(2012)]{Costigan12} Costigan, G.; Scholz, A.; Stelzer, B. et al. 2012, MNRAS, 427, 1344
\bibitem[Dahm(2008)]{Dahm08} Dahm, S.E. 2008, AJ, 136, 547
\bibitem[Donehew \& Brittain(2011)]{Donbrit11} Donehew, B., \& Brittain, S. 2011, AJ, 141, 46
\bibitem[Dullemond et al.(2006)]{Dullemond06} Dullemond C.D.; Natta A.; Testi L. 2006, ApJ, 645, L69
\bibitem[Dupree et al.(2012)]{Dupree12} Dupree, A.K.; Brickhouse, N.S.; Cranmer, S.R. et al. 2012, ApJ, 750, 73
\bibitem[Edwards et al.(2006)]{Edwards06} Edwards, S.; Fischer, W.; Hillenbrand, L.; Kwan, J. 2006, ApJ, 646, 319
\bibitem[Ercolano et al.(2014)]{Ercolano14} Ercolano, B.; Mayr, D.; Owen, J.E.; Rosotti, G.; Manara, C.F. 2014, MNRAS, 439, 256
\bibitem[Fairlamb et al.(2015)]{Fairlamb15} Fairlamb J.R.; Oudmaijer, R.D.; Mendigut\'{\i}a; I., Ilee; J.D., van den Ancker, M.E. 2015, MNRAS, accepted.
\bibitem[Fang et al.(2009)]{Fang09} Fang, M.; van Boekel, R.; Wang, W. et al. 2009, A\&A, 504, 461 
\bibitem[Hartigan et al.(1995)]{Hartigan95} Hartigan, P.; Edwards, S.; Ghandour, L. 1995, ApJ, 452, 736
\bibitem[Herczeg \& Hillenbrand(2008)]{HerczegHillen08} Herczeg, G.J.; Hillenbrand, L.A. 2008, ApJ, 681, 594
\bibitem[Kenyon \& Hartmann(1995)]{KennyonHartmann95} Kenyon, S.J. \& Hartmann, L. 1995, ApJS, 101, 117
\bibitem[Kurosawa \& Romanova(2012)]{Kurosawa12} Kurosawa, R.; Romanova, M.M. 2012 2012, MNRAS, 426, 2901 
\bibitem[Kurosawa et al.(2011)]{Kurosawa11} Kurosawa, R.; Romanova, M.M.; Harries, T.J. 2011, MNRAS.416, 2623
\bibitem[Kurosawa et al.(2006)]{Kurosawa06} Kurosawa, R.; Harries, T.J.; Symington, N.H. 2006, MNRAS, 370, 580 
\bibitem[Kurucz(1993)]{Kurucz93} Kurucz, R. L. 1993, synthe Spectrum Synthesis Programs and Line Data, Kurucz CD-ROM (Cambridge, MA: Smithsonian Astrophysical Observatory)
\bibitem[Manara et al.(2013)]{Manara13} Manara, C.F.; Testi, L.; Rigliaco, E. et al. 2013, A\&A, 551, A107
\bibitem[Meeus et al.(2012)]{Meeus12} Meeus, G.; Montesinos, B.; Mendigut\'{\i}a, I. et al. 2012, A\&A, 544, A78
\bibitem[Mendigut\'{\i}a et al.(2011)]{Mendi11} Mendigut\'{\i}a, I., Calvet, N., Montesinos, B. et al. 2011, A\&A, 535, A99
\bibitem[Mendigut\'{\i}a et al.(2013a)]{Mendi13a} Mendigut\'{\i}a, I., Brittain, S., Eiroa, C. et al. 2013, ApJ, 776, 44
\bibitem[Mendigut\'{\i}a(2013b)]{Mendi13b} Mendigut\'{\i}a, I. 2013, AN, 334, 129
\bibitem[Mohanty et al.(2005)]{Mohanty05} Mohanty, S.; Jayawardhana, R.; Basri, G. 2005, ApJ, 626, 498
\bibitem[Muzerolle et al.(2004)]{Muzerolle04} Muzerolle, J.; D'Alessio, P.; Calvet, N.; Hartmann, L. 2004, ApJ, 617, 406
\bibitem[Muzerolle et al.(1998a)]{Muzerolle98a} Muzerolle, J.; Calvet, N.; Hartmann, L. 1998a, ApJ, 492, 743
\bibitem[Muzerolle et al.(1998b)]{Muzerolle98b} Muzerolle, J.; Hartmann, L.; Calvet, N. 1998b, AJ, 116, 455
\bibitem[Muzerolle et al.(1998c)]{Muzerolle98c} Muzerolle, J.; Hartmann, L.; Calvet, N. 1998c, AJ, 116, 2965
\bibitem[Natta et al.(2014)]{Natta14} Natta, A.; Testi, L.; Alcal\'a, J.M. et al. 2014, A\&A, 569, A5
\bibitem[Natta et al.(2006)]{Natta06} Natta, A.; Testi, L.; Randich, S. 2006, A\&A, 452, 245
\bibitem[Nguyen et al.(2009)]{Nguyen09} Nguyen, D.C.; Scholz, A.; van Kerkwijk, M.H.; Jayawardhana, R.; Brandeker, A. 2009, ApJL, 694, L153
\bibitem[Padoan et al.(2005)]{Padoan05} Padoan P.; Kritsuk A.; Norman M.; Nordlund, $\AA{}$. 2005, ApJ, 622, L61
\bibitem[Rigliaco et al.(2015)]{Rigliaco15} Rigliaco, E.; Pascucci, I.; Duchene, G. et al. 2015, ApJ, complete.
\bibitem[Rigliaco et al.(2012)]{Rigliaco12} Rigliaco, E.; Natta, A.; Testi, L. et al. 2012, A\&A, 548, A56
\bibitem[Rigliaco et al.(2011)]{Rigliaco11} Rigliaco, E.; Natta, A.; Randich, S. et al. 2011, A\&A, 526, L6
\bibitem[Tilling et al.(2008)]{Tilling08} Tilling, I., Clarke, C.J., Pringle, J.E., Tout, C.A. 2008, MNRAS, 385, 1530
\bibitem[Vorobyov \& Basu(2008)]{VorobyovBasu08} Vorobyov, E.I. \& Basu, S. 2008, ApJ, 676, L139 
\bibitem[Zhou et al.(2014)]{Zhou14} Zhou, Y.; Herczeg1, G.J.; Kraus, A.L.; Metchev, S.; Cruz, K.L. 2014, ApJL, 783, 1
\end{thebibliography}
\end{document}